\documentclass[]{spie}  

\usepackage{amsmath,amsfonts,amssymb}
\usepackage{graphicx}
\usepackage[colorlinks=true, allcolors=blue]{hyperref}

\usepackage{array}
\newcolumntype{$}{>{\global\let\currentrowstyle\relax}}
\newcolumntype{^}{>{\currentrowstyle}}
\newcommand{\rowstyle}[1]{\gdef\currentrowstyle{#1}%
  #1\ignorespaces
}

\title{In-flight performance of the NIRSpec Micro Shutter Array}

\author[a]{Timothy~D.~Rawle}
\author[b]{Giovanna~Giardino}
\author[c]{David~E.~Franz}
\author[c]{Robert~Rapp}
\author[a]{Maurice~te~Plate}
\author[c]{Christian~A.~Zincke}
\author[d]{Yasin~M.~Abul-Huda}
\author[e]{Catarina~Alves~de~Oliveira}
\author[d]{Katie~Bechtold}
\author[d]{Tracy~Beck}
\author[a]{Stephan~M.~Birkmann}
\author[a]{Torsten~B\"oker}
\author[f]{Ralf~Ehrenwinkler}
\author[e]{Pierre~Ferruit}
\author[d]{Dennis~Garland}
\author[g]{Peter~Jakobsen}
\author[d]{Diane~Karakla}
\author[f]{Hermann~Karl}
\author[d]{Charles~D.~Keyes}
\author[d]{Robert~Koehler}
\author[h]{Nimisha~Kumari}
\author[a]{Nora~L\"utzgendorf}
\author[h]{Elena~Manjavacas}
\author[e]{Anthony~Marston}
\author[i]{S.~Harvey~Moseley}
\author[f]{Peter~Mosner}
\author[d]{James~Muzerolle}
\author[d]{Patrick~Ogle}
\author[d]{Charles~Proffitt}
\author[d]{Elena~Sabbi}
\author[a]{Marco~Sirianni}
\author[d]{Glenn~Wahlgren}
\author[d]{Emily~Wislowski}
\author[j]{Raymond~H.~Wright}
\author[d]{Chi~Rai~Wu}
\author[h]{Peter~Zeidler}
\affil[a]{European Space Agency (ESA), ESA Office, STScI, Baltimore, MD 21218, USA}
\affil[b]{ATG Europe for European Space Agency (ESA), ESTEC, Noordwijk, Netherlands}
\affil[c]{NASA Goddard Space Flight Center, Greenbelt, MD 20771, USA}
\affil[d]{Space Telescope Science Institute (STScI), Baltimore, MD 21218, USA}
\affil[e]{European Space Agency (ESA), ESAC, 28692 Villanueva de la Cañada, Madrid, Spain}
\affil[f]{Airbus Defence and Space GmbH, Ottobrunn, Germany}
\affil[g]{Cosmic Dawn Center, Niels Bohr Institute, University of Copenhagen, Copenhagen, Denmark}
\affil[h]{AURA for European Space Agency (ESA), ESA Office, STScI, Baltimore, MD 21218, USA}
\affil[i]{Quantum Circuits, Inc., New Haven, CT, USA}
\affil[j]{Ball Aerospace, Boulder, CO, USA }

\authorinfo{T.D.R.: E-mail:  tim.rawle@esa.int}

\begin{document} 
\maketitle

\begin{abstract}

The NIRSpec instrument on the James Webb Space Telescope (JWST) brings the first multi-object spectrograph (MOS) into space, enabled by a programmable Micro Shutter Array (MSA) of $\sim$250,000 individual apertures. During the 6-month Commissioning period, the MSA performed admirably, completing $\sim$800 reconfigurations with an average success rate of $\sim$96\% for commanding shutters open in science-like patterns. We show that 82.5\% of the unvignetted shutter population is usable for science, with electrical short masking now the primary cause of inoperable apertures. In response, we propose a plan to recheck existing shorts during nominal operations, which is expected to reduce the number of affected shutters. We also present a full assessment of the Failed Open and Failed Closed shutter populations, which both show a marginal increase in line with predictions from ground testing. We suggest an amendment to the Failed Closed shutter flagging scheme to improve flexibility for MSA configuration planning. Overall, the NIRSpec MSA performed very well during Commissioning, and the MOS mode was declared ready for science operations on schedule.
\end{abstract}


\section{Introduction}
\label{sec:intro} 

NIRSpec is the primary dedicated near-infrared (0.6--5.3$\mu$m) spectrograph on the James Webb Space Telescope (JWST), launched on 25 December 2021. This versatile instrument\cite{2022A&A...661A..80J} provides multiple complementary capabilities to explore a wide range of science themes. Users are given a choice in wavelength range and spectral resolution (R$\sim$100 over the full range, or R$\sim$1000, 2700 in one of three wavelength bands) as well as several aperture options: precision fixed slits for high-contrast high-throughput single-object spectroscopy, such as in time-series observations \cite{2022A&A...661A..83B}, a 3$\times$3 arcsec integral field unit (IFU) for spatially-resolved spectroscopy \cite{2022A&A...661A..82B}, and, as a first for an observatory in space, the Micro Shutter Array (MSA) enabling multi-object spectroscopy (MOS) \cite{2022A&A...661A..81F}. All of the NIRSpec modes were approved for science operations by the end of the six-month JWST Commissioning phase \cite{2022SPIE...torsten, 2022arXiv220705632R}, with the MOS mode declared ready on 1~July~2022.

These proceedings are concerned with the operability of the MSA hardware\cite{2008SPIE.7010E..3DK} for the MOS mode. The MSA comprises four quadrants, each housing 365$\times$171 shutters giving a total of 249,660 possible apertures. Each can be opened and closed individually, enabling the user to create a custom slit-mask to select tens to several hundred targets for simultaneous observation, while blocking background contamination from the rest of the 3.2$\times$3.4 arcmin field of view.

Every shutter of the micro-electro-mechanical system (MEMS) MSA is a 78$\times$178$\mu$m door, containing an embedded electrode and etched by magnetic strips, which opens into a crate-like housing with a second electrode on the hinge-side wall. Reconfiguration of the shutter array is aided by a Magnet Arm (MagArm) that traverses across the MSA between Primary Park on one side and Secondary Park on the other. While at Primary Park, a semi-circular tab mounted on the arm obscures the IFU aperture, so for IFU observations the MagArm is moved to a third location adjacent to Primary Park but slightly further from the array to unblock the IFU aperture. MSA shutter reconfiguration occurs in two steps: 1) all shutters are opened into their crates by sweeping the MagArm from Primary Park across the array to Secondary Park, while simultaneously setting the door and crate electrodes to opposite potentials; 2) the MagArm returns across the array to Primary Park while, column-by-column\footnote{These proceedings follow standard orientation for detectors and the MSA, and hence slitlets are opened in shutter ``columns" perpendicular to the MagArm direction of motion while ``rows" correspond to the dispersion direction: each quadrant comprises 365 columns of 171 shutters.}, each shutter in the column is simultaneously and individually addressed, maintaining a potential (electrostatic charge) to latch those commanded to open, while pulsing a zero potential to the others allowing them to close. Each double-pass sweep of the MagArm takes a little over 25 seconds. All NIRSpec exposures are executed while the MagArm is in either Primary Park or IFU position.

When assessing MSA operability, and the resulting impact on science observations, it is worth noting that for MOS, known stable failures can be accounted for during programme preparation, allowing users to avoid contamination of target traces albeit at the expense of reduced multiplexing (i.e. fewer simultaneous objects). In contrast, newly appeared or transient failures are likely to disrupt or degrade a fraction of the MOS science data. Unfortunately, some modes of MSA shutter failure (particularly Failed Open, see Section \ref{sec:FO}) impact IFU observations, as spectra from the pseudo-slits fall on the same detector pixels as the shutter traces (c.f. the fixed slits illuminate a separate detector area). Indeed, as IFU traces are always incident on the same pixels (for a given dispersive element) contamination is unavoidable, although observational techniques such as IFU-closed leakage exposures \cite{2018SPIE10698E..5ND} and dithering can mitigate most of the impact.

\section{In-flight operations}
\label{sec:flight}

The MSA operability assessment presented in these proceedings refer to the in-flight Commissioning (COMM) period from launch on 25 December 2021 until the final NIRSpec commissioning activity on 28 June 2022.

For long-term trending, data are incorporated from two ground test campaigns: Cryo-Vacuum \#3 (CV3) at NASA Goddard Space Flight Center, Maryland in the winter of 2015--16 \cite{2016SPIE.9904E..0BB}, and the Optical Telescope element / Integrated Science instrument module (OTIS) test at NASA Johnson Space Center, Texas in summer 2017 \cite{2018SPIE10698E..07T, 2018SPIE10698E..3QR}. Data from earlier ground tests are excluded as the current Flight Module MSA was only installed in NIRSpec in mid-2015. Note that operation of the MagArm, and hence MSA, strictly requires a cryo-vacuum environment.

\subsection{MSA reconfigurations}
\label{sec:reconfigs}

On 11 February 2022, the MSA temperature dropped into the allowable operational range for the MagArm ($<$ 72 K), so the MSA Control Electronics (MCE) were initialised and the Launch Lock was successfully released at the first attempt, moving the MagArm from stowed into Primary Park. The first in-flight sweep of the MagArm across the shutters occurred on 27 February 2022, once the full optical bench reach safe operating temperature for various mechanism ($<$ 42 K) \cite{2022SPIE...torsten}.

In total, 798 reconfigurations were performed over 121 days in Commissioning, 6.6 per day on average. However, the actual cadence was not uniform, with the most intense period occurring between mid-April and the end of May: 594 reconfigurations over 46 days (12.9 per day). For comparison CV3 comprised 459 over 65 days (7.1 per day) and OTIS had 200 over 42 days (4.8 per day).

Standard MSA configurations were used extensively throughout Commissioning, including the self-descriptive ALLOPEN and ALLCLOSED and the pair of full checkerboards (CHKBD1x1-n, n=1,2), used primarily for shutter operability assessment. Instrument model registration employs a pattern of crosses (CROSS5) and a sparse checkerboard which opens every 5th shutter on every other line (CHKBD3x3), while full model fitting uses dashed slitlets scattered sparsely across the MSA \cite{2022SPIE...nora, 2022SPIE...catarina}. Spectroscopic flats for the MOS mode are generated from a set of long-slits \cite{2016SPIE.9904E..46R}, opening full columns in two or four quadrants. Finally, configurations for external astrometric calibration replicate the expected science usage, tightly packing 3-shutter slitlets over the full MSA.

\section{Shutter operability}
\label{sec:MSOP}

Pre-flight testing of the MSA uncovered a number of modes by which individual shutters could be rendered unusable for science operations. The following subsections detail each of these in turn, describing the in-flight trends and comparing to ground test data \cite{2016SPIE.9904E..0BB, 2018SPIE10698E..3QR, 2022A&A...661A..81F}.

\subsection{Vignetting}
\label{sec:VIG}

The NIRSpec Field Stop, located at the telescope focus in the instrument foreoptics, is intentionally undersized compared to the field of view covered by MSA shutters. For on-sky exposures, shutters on the outer edges of each quadrant will not be illuminated and therefore cannot be used as science apertures. For internal lamp observations, the foreoptics are not part of the optical path, so they are clear of vignetting by the Field Stop, although the three upper- and lowermost rows actually project onto the focal plane beyond the extent of the detector pixels. During CV3 all exposures were internal, and even at OTIS, external observations were limited in number and only viewed point sources offering little illumination of the full field. Therefore, the pre-launch map of vignetted shutters was estimated by combining earlier instrument-level exposure data with direct photography of the MSA as seen on the JWST tertiary mirror, performed after integration of the instrument module and telescope. The total number of affected shutters is presented in Table~\ref{tab:vig}.

\begin{figure}
\centering
\includegraphics[width=17.4cm]{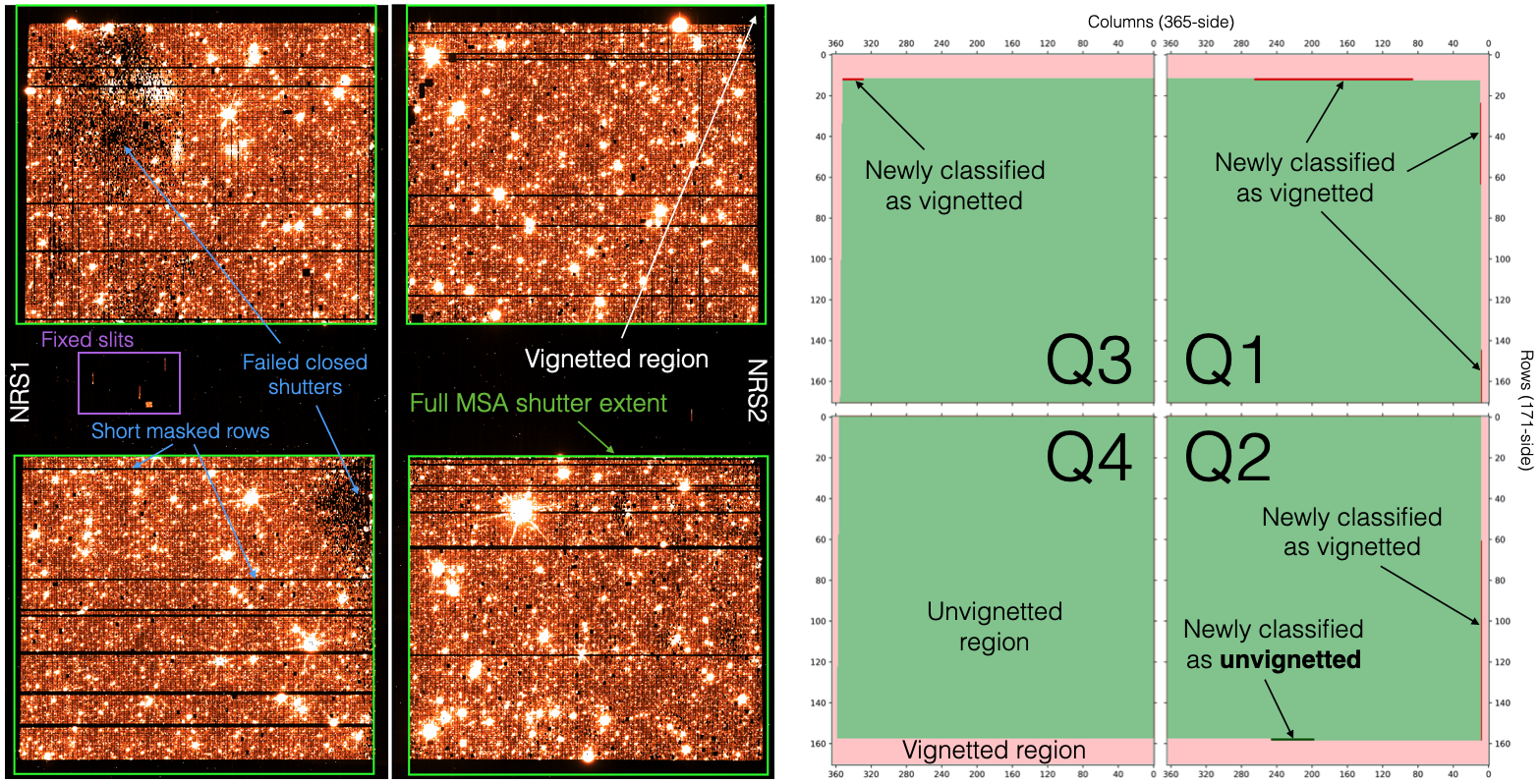}
\caption{\textbf{Left panel:} Imaging count-rate map of the JWST astrometric field (configuration = ALLOPEN, filter = F140X, exposure time = 354 s). The full extent of each MSA quadrant is indicated by a green box. Labels identify vignetting, as well as a few of the short-masked rows and areas of many Failed Closed shutters (see later sections). Detector regions beyond the MSA imprint have been cropped, as has the detector gap, but the fixed slit region between upper and lower quadrants remains. \textbf{Right panel:} Diagram highlighting the change in vignetting classification from pre-launch to flight: light green (most shutters) = not vignetted, pink = vignetted, red = newly reclassified as vignetted, dark green = now considered unvignetted. The orientation of the MSA is identical in both panels.}
\label{fig:vig}
\end{figure}

In flight, the first external observations by NIRSpec were executed on 5~April~2022 during the early-checkout astrometric calibration programme (PID1120), point-and-shoot imaging of a crowded stellar field in the Large Magellanic Cloud. The fixed pattern of shadows from short-masked and plugged shutters (evident in the left panel, Figure~\ref{fig:vig}) helped to register the MSA grid to the detector pixels, thereby enabling identification of shutters shadowed by the Field Stop. However, bright objects at the edge of the unvignetted area, together with the variable background of the field, ensure that a binary identification of vignetted shutters is non-trivial. A conservative classification was adopted, resulting in a marginal increase in the number of shutters considered to be vignetted (Table~\ref{tab:vig} and right panel, Figure~\ref{fig:vig}), but there is no evidence of a physical shift of the Field Stop relative to the MSA plane. Further external imaging on 28~April confirmed the initial analysis.

\begin{table}
\caption{Number of shutters vignetted by the NIRSpec Field Stop, estimated at ground testing and then re-derived using on-sky data during Commissioning (COMM). The final column quantifies the vignetted population as a percentage of the total number of shutters.}
\centering
\begin{tabular}{llcccccr}
\\
\textbf{Vignetted} & & \textbf{Q1} & \textbf{Q2} & \textbf{Q3} & \textbf{Q4} & \textbf{Total} & \textbf{\%} \\
\hline
Ground testing & 2015--2017 & 5873  & 5880 &  6078  & 5874  &  23705  &   9.5\% \\
COMM & 2022 & 6119  & 5929  & 6102  & 5874  &  24024  &   9.6\% \\
\end{tabular}
\label{tab:vig}
\end{table}

\subsection{Electrical shorts and masking}
\label{sec:SM}

Particulate contamination in the complex MCE circuitry can cause electrical shorts leading to unsafe currents, bright thermal emission contaminating the detectors and/or unpredictable shutter behaviour that could degrade science data. In most cases, a new short is discovered by passive monitoring of instrument telemetry. Specifically, a new short will manifest as an elevated quadrant return-rail current while a configuration is held on the array (current spikes are expected during MagArm sweeps irrespective of shorts). The short can be located via a dedicated ``electrical short detection" procedure that sequentially varies line voltages while automatically monitoring for changes in current, thereby executing without physically moving the limited-life MagArm.

\begin{figure}
\centering
\includegraphics[width=17cm]{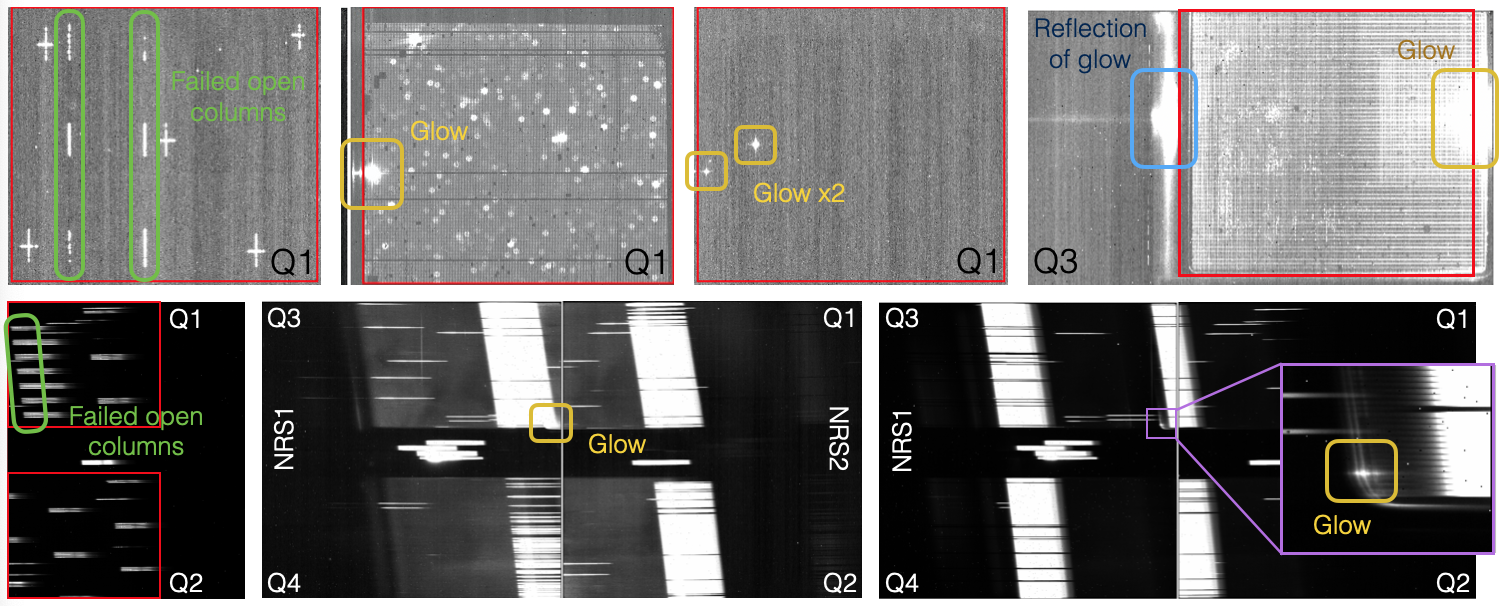}
\caption{Impact of shorts as seen by the NIRSpec detectors. \textbf{Upper row:} The effects of a short are easier to identify in imaging mode, but that will be used sparingly during normal operations. \textit{Left--right:} Short-induced Failed Open columns mimic the commanded open rows from other columns (internal lamp exposure); bright glow from a short has distinctively different diffraction spikes compared to on-sky point sources (4 rather than 6); glow from two shorts in a dark exposure; strong shorts can drown out all other features on the detector, even causing internal reflections -- this short was also detected in telemetry. \textbf{Lower row:} The impact of shorts are more difficult to spot in dispersed observations. \textit{Left:} Induced Failed Open columns are buried amongst commanded dashed slits in this internal prism exposure. As above, the failure mimics the location of other open shutters in the quadrant. A careful examination of the density of traces belies the presence of the short, but actually, due to the sparse pattern, the data from this exposure suffers no ill effects from the short. \textit{Centre and right:} Internal prism exposures through two different long-slits. Low-level glow is present in both, contaminating the traces in the centre panel. The zoom-in shows that the short illuminates the rail on the MSA edge.\\}
\label{fig:shorts}
\end{figure}

Low-level shorts may not elevate currents above the noise-limited threshold of the instrument telemetry, but can still generate anomalies visible in exposure data. Several examples are shown in Figure~\ref{fig:shorts}, including direct glow from the short and induced shutter failure. As low-level shorts cannot be located via the electrical detection routine, the offending row and/or column must be deduced from the position of the anomalies, either directly from existing affected data, or via dedicated ``optical short detection" exposures.\cite{2018SPIE10698E..3QR}

Whether detected electrically or optically, all shorts are mitigated by masking an entire shutter row or column directly in the MCE. As a masked column only removes 171 shutters from operation, compared to 365 for a row, the former are chosen preferentially if possible.

Table~\ref{tab:sds} lists all executed short detections during CV3, OTIS and Commissioning, detailing how many rows and columns each add to the mask, which generally corresponds to the number of new individual shorts present. During Commissioning, 15 rows and columns were added to the short mask, with new shorts appearing every $\sim$150–200 reconfigurations. Note that the 3 events in May occurred during intense MSA activity, with an average of 1 sweep per hour for 20 days, culminating in 100 reconfigurations on 30~May. This is well beyond the expected usage of normal operations, in which single reconfigurations will often be separated by long (multi-hour) science exposures. The final short of Commissioning was truly transient, only appearing for one reconfiguration, before disappearing again in subsequent repeats of the same pattern. As documented in Table~\ref{tab:sm}, each masked row/column impacts a large number of shutters, with $\sim$3200 additional shutters succumbing to the mask in Commissioning.

\begin{table}
\caption{Timeline of short detection during CV3, OTIS and Commissioning (COMM), showing the number of rows/columns each episode added to the mask.}
\centering
\begin{tabular}{$l^l^r^c}
\\
\multicolumn{1}{l}{\textbf{Date}} & \multicolumn{1}{l}{\textbf{Description}} & \multicolumn{2}{c}{\textbf{Added rows/columns}} \\
\hline
\multicolumn{4}{l}{\textbf{CV3}} \\
23 Nov 2015 & Electrical \#1 & 6 & \\
30 Nov 2015 & Electrical \#2 & 5 & \\
30 Nov 2015 & Optical \#1 & 2 & \\
09 Dec 2015 & Electrical \#3 & 2 & \\
03 Jan 2015 & Electrical \#4 & 2 & \\
06 Jan 2015 & Optical \#2 & 1 & \\
\multicolumn{2}{l}{\textbf{Total CV3}} & & \textbf{18} \\
\hline
\multicolumn{4}{l}{\textbf{OTIS}} \\
13 Aug 2017 & Electrical \#1 & 3 & \\
21 Aug 2017 & Optical \#1 & 1 & \\
02 Sep 2017 & Electrical \#2 & 3 & \\
09 Sep 2017 & Electrical \#3 & 0 & \\
11 Sep 2017 & Optical \#2 & 1 & \\
17 Sep 2017 & Optical \#3 & 1 & \\
\multicolumn{2}{l}{\textbf{Total OTIS}} & & \textbf{9} \\
\hline
\multicolumn{4}{l}{\textbf{COMM}} \\
27 Feb 2022 & Electrical \#1 & 2 & \\
09 Mar 2022 & Electrical \#2 & 1 & \\
23 Mar 2022 & Electrical \#3 & 2 & \\
26 Mar 2022 & Optical \#1 & 5 & \\
14 May 2022 & Optical \#2 & 1 & \\
20 May 2022 & Electrical \#4 & 1 & \\
02 Jun 2022 & Optical \#3 & 2 & \\
22 Jun 2022 & Electrical \#5 & 1 & \\
23 Jun 2022 & Electrical \#6 & 0 & \\
\multicolumn{2}{l}{\textbf{Total COMM}} & & \textbf{15} \\
\end{tabular}
\label{tab:sds}
\end{table}

\begin{table}
\caption{Detailed statistics for the initial and final short masks of CV, OTIS and Commissioning (COMM), including the change between each snapshot. Tabulated values include both the total number of masked rows/columns per quadrant (Q1--Q4, Total) and the number (and \%) of unvignetted shutters removed by the mask. CV3--OTIS: re-examination of the data led to two lines being removed from the mask. OTIS--COMM: the short mask remained unaltered and the slight decrease in shutters here arises from the change in identified vignetted shutters.}
\centering
\begin{tabular}{$l^r^r^r^r^r^r^r}
\\
 & \multicolumn{5}{c}{\textbf{Masked rows/columns}} & \multicolumn{2}{c}{\textbf{Masked shutters}} \\
 & \multicolumn{1}{c}{Q1} & \multicolumn{1}{c}{Q2} & \multicolumn{1}{c}{Q3} & \multicolumn{1}{c}{Q4} & \multicolumn{1}{c}{Total} & \multicolumn{1}{c}{\#} & \multicolumn{1}{c}{\%} \\
\hline
\rowstyle{\bfseries}
Start CV3 & 20 & 14 & 10 & 25 & 69 & 14342 & 6.3\% \\
$\Delta$ CV3& +9 & +2 & +3 & +4 & +18 & +4089 & +1.9\% \\
\rowstyle{\bfseries}
End CV3 & 29 & 16 & 13 & 29 & 87 & 18431 & 8.2\% \\
$\Delta$ CV3--OTIS & --2 & 0 & 0 & 0 & --2 & --308 & --0.2\% \\
\rowstyle{\bfseries}
Start OTIS & 27 & 16 & 13 & 29 & 85 & 18123 & 8.0\% \\
$\Delta$ OTIS& +2 & +5 & +2 & 0 & +9 & +2327 & +1.1\% \\
\rowstyle{\bfseries}
End OTIS & 29 & 21 & 15 & 29 & 94 & 20450 & 9.1\% \\
$\Delta$ OTIS--COMM & 0 & 0 & 0 & 0 & 0 & --10 & --0.0\% \\
\rowstyle{\bfseries}
Start COMM & 29 & 21 & 15 & 29 & 94 & 20440 & 9.1\% \\
$\Delta$ COMM& +12 & +1 & +2 & 0 & +15 & +3188 & +1.4\% \\
\rowstyle{\bfseries}
End COMM & 41 & 22 & 17 & 29 & 109 & 23628 & 10.5\% \\
\end{tabular}
\label{tab:sm}
\end{table}

\subsection{Failed Open shutters}
\label{sec:FO}

Failed Open (FO) shutters do not close when commanded to release, usually because the door is warped and sticks. The acoustic load of the simulated (during ground testing) and actual launch environment is usually considered to be the primary contributor for this failure mode. The FO shutters cause an unwanted trace on the detector, contaminating any overlapping target spectrum, not only in MOS mode but also for the IFU. As this impact is potentially very inhibitive for science data, any FO shutter identified during component-level testing was permanently plugged and turned into a Failed Closed shutter instead.

Remaining and subsequent FO shutters are mitigated by flagging the affected pixels at the planning stage, so targets that would illuminate the same pixels can be avoided. As such, the FO population directly impacts overall multiplexing, with the loss of approximately 1.5 potential targets for every FO shutter, depending on both the dispersive element and the input source catalogue density. Several observed FO shutters are actually only partially open, with a much lower throughput than fully open shutters. Due to their uncertain slit loss characteristics, users are advised not to risk using any FO shutters for science.

\begin{table}
\caption{Failed Open (FO) shutters during CV3, OTIS and at two epochs of Commissioning (COMM). For consistency, these numbers exclude vignetted FO shutters.}
\centering
\begin{tabular}{lrccccc}
\\
\textbf{FO} & & \textbf{Q1} & \textbf{Q2} & \textbf{Q3} & \textbf{Q4} & \textbf{Total} \\
\hline
Final CV3 & 2015 & 5 & 3 & 9 & 2 & 19 \\
Final OTIS & 2017 & 4 & 3 & 10 & 1 & 18 \\
Initial COMM & 2022 & 6 & 3 & 10 & 1 & 20 \\
Final COMM & 2022 & 6 & 3 & 12 & 1 & 22 \\
\end{tabular}
\label{tab:fo}
\end{table}

Previously, ALLCLOSED exposures were considered to give the most complete assessment of FO shutters as they include no commanded open shutters. A blob-finding algorithm is applied to detector data after cosmic ray and snowball rejection \cite{2022SPIE...stephan}, with the pixel coordinates of candidates fed through the instrument model transforms to obtain the location in shutter-space. Fortunately, the number of FO shutters is small. Indeed no new FO shutter appeared while the MSA was in the cryo-vacuum chamber during ground testing, and one actually disappeared. As Table~\ref{tab:fo} presents, the population of FO shutters had increased by two  when derived for the first time in Commissioning (from a combination of two ALLCLOSED exposures), and this was attributed to the launch. Unfortunately, the last ALLCLOSED observations of commissioning uncovered a further two FO shutters and these were added to the initial count for the final mask. The newest FOs were assumed to be due to general wear during the intense calibration period.

\begin{figure}
\centering
\includegraphics[width=17.8cm]{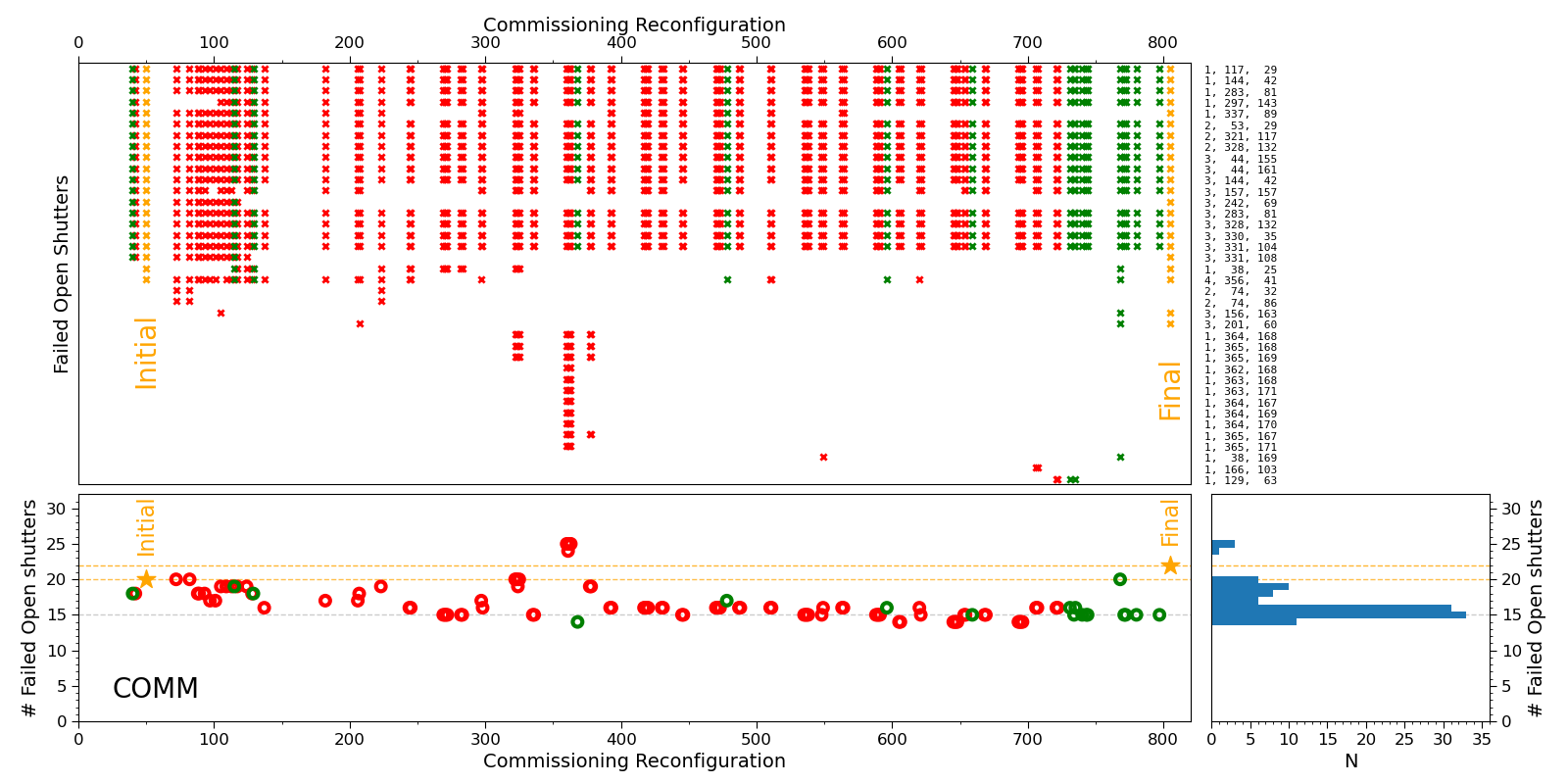}
\caption{\textbf{Upper panel:} Unvignetted Failed Open (FO) shutters for each ``primarily-closed" (not checkerboard or ALLOPEN) reconfiguration in Commissioning, displayed chronologically left to right. For every Failed Open shutter, identified on the right (quadrant, column, row), each failure is marked by a cross: orange = primary assessment as displayed in Table~\ref{tab:fo}, green = ALLCLOSED, red = other patterns. \textbf{Lower panels:} Tally of the number of FO shutters per reconfiguration, and histogram showing the total distribution. Horizontal dashed lines show the initial and final counts (orange; also marked by labelled stars) and the distribution mode (grey).}
\label{fig:fo}
\end{figure}

However, a more complete picture is achievable using all ``primarily-closed" (i.e. not checkerboard or ALLOPEN) imaging exposures. Presented in Figure~\ref{fig:fo}, this reveals a more complicated, but ultimately more positive, story than the initial and final counts alone. There are a core of 15 unvignetted shutters failing to close in almost every reconfiguration. A further 5 shutters failed consistently at the beginning of Commissioning, but two of those have not been seen again since reconfiguration $\sim$125 and another disappeared at around $\sim$570. The other two in the initial count are seen intermittently throughout Commissioning, although they are known to be only partially open, so it is difficult to confirm whether this is due to inconsistent detection of the low-level throughput or true transient failure. The two new FO shutters in the final mask are very sporadic, but both first appeared relatively soon after launch, contradicting wear as the cause -- they could also be due to the launch environment, but less severely warped and so more intermittent. It is worth mentioning that a few other shutters also failed once or twice in flight, and while they were not picked up by the baseline assessment, they are included in Figure~\ref{fig:fo}. Full Failed Open columns caused by shorts (Section \ref{sec:SM}) have been excluded from this analysis, but Figure~\ref{fig:fo} does show one long-slit pattern (reconfigured several times) that triggered a cluster of FO shutters on the edge of Q1.

The conservative final FO shutter list from Commissioning, as presented in Table~\ref{tab:fo} and labelled in Figure~\ref{fig:fo}, was delivered to science users for Cycle 1 planning. Detailed analysis will continue during nominal operations, and may indicate that the most transient pose little risk to science data and therefore may be removed from the FO shutter list.

\subsection{Failed Closed shutters}
\label{sec:FC}

Failed Closed (FC) shutters are those that either stick closed when latching, or fail to hold open when being addressed. As described in the previous section, a few Failed Open shutters were replaced by static plugs during component level testing. This failure mode is also believed to be due to warped and jammed shutters.

During science preparation, MSA configuration planning currently treats FC shutters as indistinguishable from shutters rendered unusable by the short mask. All these apertures are simply unavailable to open by the user, although for FC shutters, the prohibition is enforced by the mask preparation software alone, rather than the flight hardware electronics. In this paradigm, as long as the population of FC shutters does not become dominant, the impact on multiplexing should not be severely disrupted.

The full Failed Closed shutter population is primarily identified from imaging exposures through pairs of CHKBD1x1 (two full, but opposite, checkerboards). Based on the commanded open shutters, the failed closed population is simply derived by probing the flux in the count-rate image at the expected location of each shutter. For consistency, this initial analysis identifies Failed Closed from the commanded-open shutters that do not intersect either the short mask or vignetting. Table~\ref{tab:fc} presents the evolution of the shutter population during CV3, OTIS and Commissioning, using the same analysis procedure on CHKBD1x1 data for each epoch. At the end of Commissioning, there is an overall FC population comprising 7.0\% of all unvignetted shutters.

\begin{table}
\caption{Failed Closed (FC) shutters during CV3, OTIS and at two epochs of Commissioning (COMM). For consistency, these numbers exclude vignetted shutters. The CV3/OTIS counts differ from those presented in earlier analysis \cite{2018SPIE10698E..3QR}, which did not fully account for short masking.}
\centering
\begin{tabular}{$l^r^r^r^r^r^r}
\\
\multicolumn{1}{l}{\textbf{FC}} & \multicolumn{1}{c}{Q1} & \multicolumn{1}{c}{Q2} & \multicolumn{1}{c}{Q3} & \multicolumn{1}{c}{Q4} & \multicolumn{1}{c}{Total} & \multicolumn{1}{c}{\% FC} \\
\hline
CV3 & 1555 &  3215 &  6516 &  1892  &  13178   &  5.8\% \\
OTIS & 1607  & 3071 &  5999 &  2748  &  13425  &   5.9\% \\
Start COMM & 1605 &  3214 &  6027 &  4143 &   14989 & 6.6\% \\
End COMM & 1569 & 3328 & 5932 & 5064 & 15893 & 7.0\% \\
\\
\\
\end{tabular}
\label{tab:fc}
\end{table}

\begin{table}
\caption{Full MSA operability report for the end of Commissioning. FO $=$ Failed Open, SM $=$ Short mask, FC $=$ Failed Closed, VIG $=$ vignetted. Percentages based on total shutter population unless indicated. Note that the separate counts for short mask rows and columns intentionally sum in excess of the SM total, as they double count each crossing point.
}
\centering
\begin{tabular}{$l^r^r^r^r^r^r^l}
\\
\multicolumn{1}{l}{} & \multicolumn{1}{c}{Q1} & \multicolumn{1}{c}{Q2} & \multicolumn{1}{c}{Q3} & \multicolumn{1}{c}{Q4} & \multicolumn{1}{c}{Total} & \multicolumn{1}{c}{\%} & \\
\hline
\textbf{Total} & 62415  &62415&  62415 & 62415  & 249660 &  100.0\% &\\
\textbf{VIG}   &    6119  & 5929 &  6102 &  5874 &   24024   &  9.6\%& \\
\hline
\textbf{FO}         &                    6   &   3   &  12   &   3    &   24   & &\\
\textbf{FO} (non-VIG)      &              6  &    3   &  12    &  1     &  22   & & \\
\hline
\textbf{SM}           &               8469 &  5391 &  3817 & 7465   & 25142  & 10.1\% & \\
... 171-side (rows) & 3285  & 3285 &  1825 &  5110 &   13505 &    5.4\% & \\
... 365-side (coloums) & 5472 &  2223  & 2052  & 2565  &  12312  &   4.9\% & \\
\textbf{SM} (non-VIG)        &        7835  & 5150  & 3466 &  7177 &   23628   & 10.5\% & of non-VIG \\
\hline
\textbf{FC}             &            3445 & 5386 &  7227  & 6605  &  22663   &  9.1\%& \\
\textbf{FC} (non-VIG)    &          1569  & 3328 &  5932 &  5064  &  15893   &  7.0\% & of non-VIG \\
\hline
\textbf{Total Failed} (FC+SM)  &   11914 & 10777 & 11044 & 14070 &   47805  &  19.1\% & \\
\textbf{Total Unfailed}  &   50501 & 51638 & 51371 & 48345 &  201855  &  80.9\% & \\
\hline
\textbf{Total Unusable} (FC+SM, non-VIG)  &   9404 & 8478 & 9398 & 12241 &   39521  &  17.5\% & of non-VIG \\
\textbf{Total Usable} (non-VIG) &   46892 & 48008 & 46915 & 44300 & 186115  &  82.5\% & of non-VIG \\
\end{tabular}
\label{tab:msop}
\end{table}

\begin{figure}
\centering
\includegraphics[width=16cm]{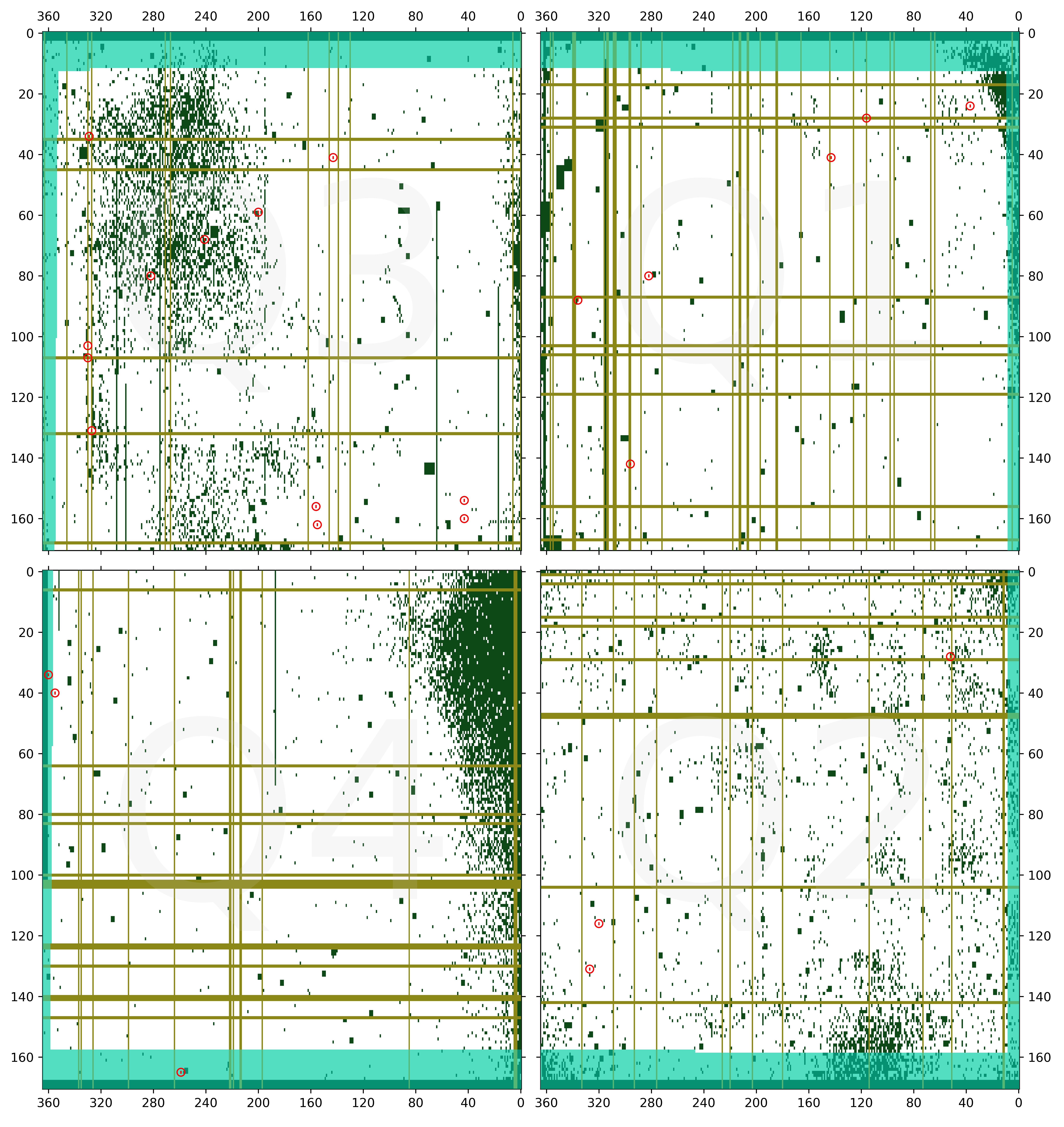}
\caption{Final Commissioning MSA operability map: dark green = Failed Closed (FC); light green = vignetted shutters (VIG); brown = short mask (SM); red = Failed Open (FO), additionally highlighted by red circles. VIG is plotted with transparency so the underlying FC and SM are visible -- note also the two vignetted FO in quadrant 4. The upper- and lowermost three rows are marked as FC, although they actually illuminate the focal plane beyond the edge of the detector. The layout and orientation of the quadrants is identical to the right panel of Figure~\ref{fig:vig}.}
\label{fig:msop}
\end{figure}

The full final Commissioning MSA operability map, including vignetting, short masks and Failed Open/Closed, is presented in Figure~\ref{fig:msop}. The detailed breakdown of the associated operability statistics is given in Table~\ref{tab:msop}. Similar regions of Failed Closed shutters have been present in Quadrants 3 and 4 since CV3, but have grown noticeable over time. As described in Section \ref{sec:FO}, Quadrant 4 also boasts the largest number of Failed Open shutters, which is unfortunate for one of the two quadrants nearest the IFU pseudo-slits. Figure~\ref{fig:msop} also highlights that Quadrant 1 has by far the most short masked rows and columns (Section \ref{sec:SM}).

\subsection{Failed Closed shutters in science-like configurations}
\label{sec:fc_usage}

Derivation of the Failed Closed population from the full checkerboard pair is known to yield an upper limit. These complex patterns, with the maximum possible number of alternating voltage pulses, produce more configuration errors: e.g. ALLOPEN configurations show as few as half the number of FC shutters compared to CHKBD1x1 at the same epoch \cite{2018SPIE10698E..3QR}. However, configurations designed for science in nominal operations will be much closer to ALLCLOSED, with only a few hundred commanded open shutters. So, although CHKBD1x1 and ALLOPEN are fine for probing the operability of every shutter in the full MSA, neither pattern necessarily gives an adequate prediction of the expected shutter failure rate for nominal operations. The question to answer is -- \textit{in a science-like configuration designed to avoid known failures, what is the chance of a commanded open shutter failing?}

This requires a more in-depth analysis of as many of the 798 Commissioning reconfigurations as possible. Of those, 146 have no associated exposures: 49 were engineering sweeps and 97 were for visit ``clean-up", executed when necessary to ensure each individually-schedulable block of exposures ends with an ALLCLOSED pattern on the MSA. Many of the first 100 reconfigurations in Commissioning were engineering sweeps. Commanded-open success also cannot be derived from the other 135 ALLCLOSED reconfigurations or the further 31 patterns used for purely dark internal exposures. The remaining 486 reconfigurations are all input for this analysis. For those with an associated imaging exposure, the flux at the detector location is probed for all commanded-open shutters, as in Section \ref{sec:FC}. For those only followed by dispersed light observations, the trace of every commanded open shutter is extracted and any without flux (i.e. closed) identified. Failed Closed quantification is less robust for on-sky observations due to the spatial variation of the flux, compared to the constant illumination of the internal lamps.

For each reconfiguration, a single figure of merit is derived: \%~success. This is calculated from the population of shutters commanded to open excluding any kept closed by the contemporaneous short mask, as that is a different mode of inoperability and science programmes should be designed around it. Note that this is subtly different from the analysis behind Table \ref{tab:fc}, as there is no reason to discount vignetted shutters for internal observations in this analysis -- vignetted shutters are generally not commanded open in configurations destined for external use, but if they were, they are ignored here, ensuring that only true Failed Closed shutters are counted. \%~success then simply quantifies those opened successfully compared to the total non-masked commanded.

\begin{figure}
\centering
\includegraphics[width=17.5cm]{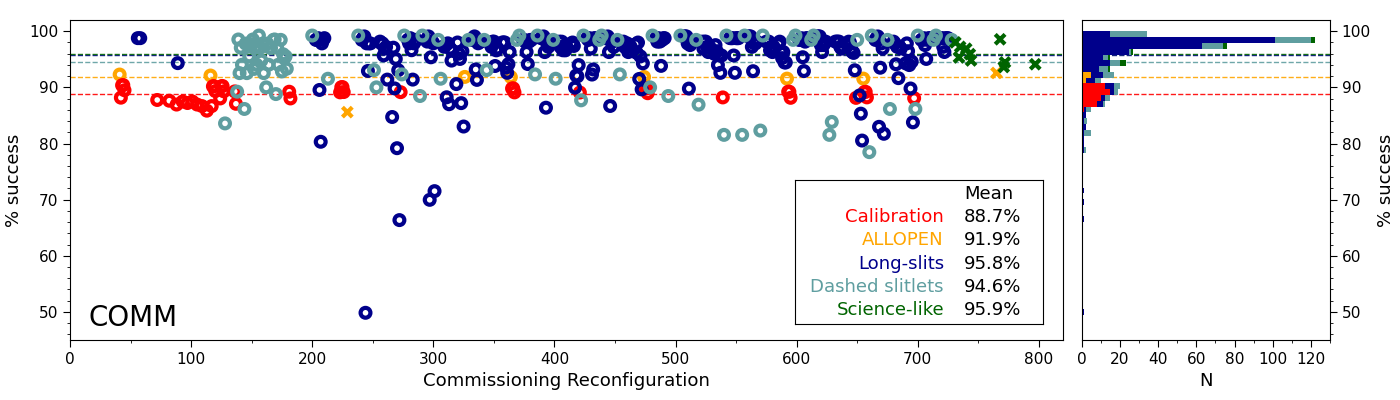}
\caption{Percentage success for commanded open shutters in each Commissioning reconfiguration, presented in chronological order. Shutters in the contemporaneous short masks are excluded. Colours distinguish broad configuration classes, described in the text. Circles denote reconfigurations analysed via an internal lamp exposure, while crosses were external only, giving less robust quantification of failure due to the variable illumination of the sky. The histogram demonstrates that science-like reconfigurations show the highest level success.}
\label{fig:fc}
\end{figure}

The full results are presented in Figure~\ref{fig:fc}, classifying similar reconfigurations together. Calibration configurations (red in Figure~\ref{fig:fc}) include checkerboards of all types and CROSS5, and are the least science-like. They are not only more complex for the MSA to address, but they are also commanded ``pristine", full patterns without trying to avoid known Failed Closed shutters, as one of their uses is to locate those failures. Their mean \%~failure of 11.3\% (c.f. 100 -- \%~success) is higher than the final Commissioning ``FC" of 9.1\% quoted in Table~\ref{tab:msop} as the initial period proved more variable in success. In contrast, but as expected, the similarly pristine ALLOPEN reconfigurations demonstrate the effect of pattern complexity with a lower mean \%~failure of 8.1\%.

Most of the long-slits (dark blue in Figure~\ref{fig:fc}) are from the spectroscopic flat-field calibration programme (PID1122)\cite{2016SPIE.9904E..46R}, which required them to be at regular intervals across the MSA. Each long-slit was nudged by up to a few columns from the baseline spacing to maximise the number of operable shutters, but most still contain multiple known FC shutters and they were all commanded pristine. PID1122 contributed 273 reconfigurations to commissioning, with 72 unique patterns each configured on 3--4 occasions. The particularly low success rate for some long-slits (as low as 50\%) was caused by the large Failed Closed regions in Quadrants 3 and 4 (see Figure~\ref{fig:msop}), which could not be fully avoided without sacrificing the required uniform spacing. Overall, however, the long-slits gained a 95.8\% success rate.

Dashed slitlets (blue-green in Figure~\ref{fig:fc}) were the first science-like configurations used by the MSA, consisting of sparsely distributed groups of 9--13 alternating open/closed shutters. Pattern design attempted to avoid known FC shutters, but, particularly earlier in Commissioning, were required to probe the quadrant corners where failure rates happen to be higher. Additionally, several dashed slitlets yielded failure rates more akin to checkerboards, raising the possibility that even short bursts of alternate open-close states may increases the likelihood of failure, although that cannot be confirmed with the data in hand. Ultimately, both these aspects reduced their success rate to just below the long-slits at 94.6\%.

Finally, several truly science-like configurations were used in external calibration programmes towards the end of commissioning (dark green crosses in Figure \ref{fig:fc}). These patterns were designed to avoid known FC shutters, and primarily comprised of standard ``three-shutter slitlets", a recommended option in the MSA planning tool \cite{2014SPIE.9149E..1ZK} employed for background subtraction via the nodding technique. As expected, these configurations proved to be the most successful with 95.9\% of commanded open shutters latching as requested. The larger variation in success rates for these configurations is due to the increased uncertainty from measurements on the sky compared to internal lamps. Although $\sim$4\% failure may sound reasonable, this is after avoiding the 7\% of unvignetted shutters already identified as Failed Closed. Table \ref{fig:fc} shows that the overall FC population grew by only 0.4\% during Commissioning, so selecting this level of as-yet-unknown permanent FC shutters is unlikely. Much more probable is a scenario in which some or even most shutters have an intrinsic probability of failure -- a transient mode of FC shutters.

\begin{figure}
\centering
\includegraphics[width=8.5cm]{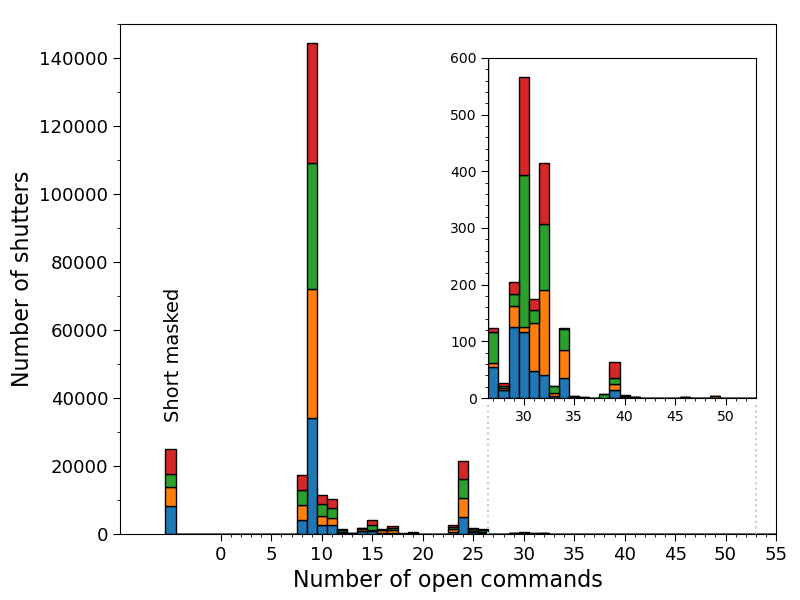}
\includegraphics[width=8.5cm]{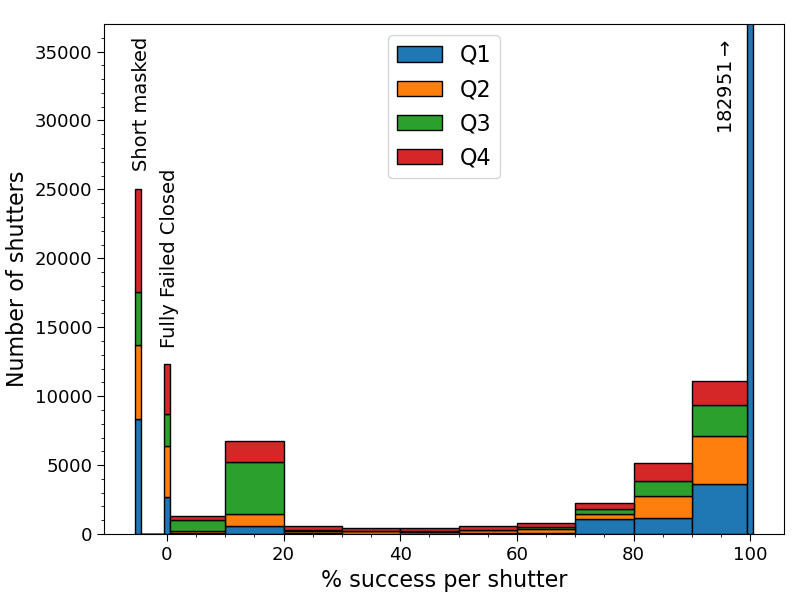}
\caption{Histograms presenting per shutter commanding (\textbf{left}) and \% success (\textbf{right}) for Commissioning reconfigurations, with quadrants coloured differently, as labelled in the right panel. Shutters in the final short mask are counted separately in both panels. The inset in the left panel highlights detail. The bar for fully successful shutters in the right panel is truncated.}
\label{fig:open}
\end{figure}

Figure~\ref{fig:open} displays the reconfiguration success at the shutter level. During Commissioning, all shutters (except those in the short mask) were commanded open at least nine times by a reconfiguration with measurable data (i.e not including darks or engineering sweeps); a handful were commanded open almost 50 times (inset of the left panel, Figure~\ref{fig:open}). The configuration history of every shutter can be tracked through Commissioning, building a full picture of shutter behaviour. The right panel of Figure~\ref{fig:open} presents the success of each shutter. The vast majority of shutters open 100\% successfully. Note that the number counted here (182,951) is lower than the comparable `Total Unfailed' value presented in Table \ref{tab:msop} (201,855), as the latter is from a single snapshot, while here only those \textbf{always} successful are included. The difference between the two is $\sim$20,000 shutters. Figure~\ref{fig:open} shows that the equivalent sized population would include shutters down to 60--70\% success rate. At the other end, the population of shutters that never responds to an open command is only 12,312 (4.9\% of all shutters), which is significantly below 9.1\% (row `FC') counted in the single reconfiguration snapshot. Even including all shutters with a success rate $<$50\%, only 8.8\% could be classified as mostly failed.

\section{Discussion and nominal operations}

\subsection{Re-checking shorts}

During Commissioning, short masking became the largest contributor to the population of unusable shutters, with 37 rows (171-side) and 72 columns (365-side) removed from active duty at the expense of more than 25,000 shutters. Q1 alone now has 31 columns masked, although the 10 masked rows of Q4 do almost as much damage to the inoperable shutter count.

At least two shorts (one each in Commissioning and OTIS) appeared and then disappeared again before a short detection could be executed. In the Commissioning example, the short failed to return in an identical reconfiguration later the same day. For the sake of efficiency, and also minimising limited life mechanism moves, programmes tend to be designed to avoid multiple reconfigurations of the same pattern close in time, raising the possibility that more shorts would be short-lived but are not given the opportunity to disappear again. Furthermore, two shorts were unmasked after CV3 and have never been seen again. This is all tantalising evidence that shorts should not be treated as permanent, a theory that is fully in line with particulate contamination as the root cause, given such particles could always dislodge or burn off and float away. 

Throughout OTIS and Commissioning, the short masking process was entirely additive, a policy determined to keep things simple during hectic periods. Starting from a clean slate would be operationally simple, but many of the shorts could still remain, and it would be potentially hazardous for the hardware to light them all at once. In reality, the procedure would involve removing one or two of the oldest shorts at a time, requiring a lot of operational resources for the flight verification and certification of  files to be uplinked to the spacecraft.

Starting in normal operations Cycle 1, the team intends to recheck previous shorts as part of the normal execution of the electrical short detection upon appearance of a new short. Rather than running the short detection on the current onboard mask, an alternate starting mask, with a few existing shorts removed, will be  preprepared and certified ready for uplink, ensuring minimal delay is incurred. The short detection then runs as normal, simply re-detecting the old shorts, if they persist, alongside the new short. In principal, as each quadrant has separate circuitry, one or two old shorts could be removed from each without any increased risk to the hardware. However, short detection runs on one quadrant at a time, and can be executed on a subset of quadrants, so testing all four quadrants instead of only one would increase the required observatory time, and such a plan would be at the discretion of mission management.

The rate of new shorts during nominal operations is not well predicted, but the hope would be to retest a significant fraction of the existing masked rows and columns over several yearly cycles, and work towards reducing the total number of masked shutters.

\subsection{Using operability in MOS preparation}

In the most basic terms, MSA planning involves the cross-matching of a user-supplied master catalogue with the shutter grid, accounting for the known operability map, in order to maximise the number of targets available for the pointed observation. The process is, of course, complicated by many further considerations \cite{2022A&A...661A..81F} not discussed here. This final section focusses on two interconnected issues directly related to shutter operability. First, the current Failed Closed population in the operability map is based on the most recent snapshot assessment, a method that assumes a constant or additively-evolving population. In the regime of transient failure, as demonstrated in Section \ref{sec:fc_usage}, this is essentially a random realisation of the underlying truth. Second, the planning software then flags all the FC shutters equally, assuming none of them are ever usable, which is also counter to the observed behaviour.

An alternate operability map would be based on the likelihood of failure for each shutters, derived from its individual history. Ideally, the planning software would then allow the user to decide what threshold of FC shutter success they are comfortable with for their science. For example, a user with a uniformly-weighted but lower-density master catalogue may simply wish to observe as many targets as possible. Multiplexing is the most important factor, and crucially the user could prefer to reduce the individual FC success threshold to obtain a superior optimisation -- for instance, one or two additional failed shutters may be acceptable, if the additional real estate means that the configuration includes more in total. On the other hand, another user may have a very small set of high priority targets, either part of a complex multi-class prioritisation scheme or simply a select few on top of a larger filler catalogue. In this case, the user may like to ensure that any included high priority targets fall into the most successful shutters. This would undoubtedly decrease overall multiplexing, but would give a much lower probability of missing the most important objects. In a  further step towards flexibility, the filler catalogue could be allowed to use a more relaxed FC success threshold than the high priority class.

Based on the Failed Closed assessment summarised in Figure \ref{fig:open}, shutters could, at minimum, be separated into 3 categories, such as: fully operational (100\% successful), FC transient (50\% $\le$ success $<$ 100\%) and FC ($<$ 50\% successful). A more sophisticated classification could try to encapsulate both \% success and the recent operational history, as it could be argued that a 70\% successful shutter that failed in each of the last 30\% of attempts is less desirable than a shutter with a random 70\% success.

The details of any improvement to the Failed Closed shutter flagging need further discussion, bolstered by an ever-increasing dataset of operability history, and certainly needs to account for the feasibility of implementation in the planning software. The analysis presented above clearly shows that a more sophisticated scheme is desirable.

\section{Summary}

NIRSpec brings the first MOS to space, enabled by a programmable Micro Shutter Array (MSA) of $\sim$250,000 apertures, allowing science users to select a few hundred targets while obscuring contamination from the rest of the $\sim$3.2x3.4 arcmin field of view. During the 6-month Commissioning period (January--June 2022), the MSA performed admirably, completing $\sim$800 reconfigurations. Overall, 82.5\% of unvignetted shutters are operable for science, with the largest contribution to the unusable population now coming from the masking of electrical shorts. We discuss a plan for rechecking existing shorts, which is expected to reduce the number of affected shutters in the future. Shutters commanded to open in science-like configurations demonstrate an average success rate of $\sim$96\%, and we emphasise that this does mean that science users should be prepared to find that $\sim$4 shutters will Fail Closed out of every hundred commanded to open, even when MOS design accounts for the known operability map. We suggest an amendment to the Failed Closed shutter flagging scheme in order to improve the clarity and flexibility of MSA configuration planning around intermittently inoperable shutters. NIRSpec MOS was declared ready for science operations on 1 July 2022.

\acknowledgments 
 
JWST is a joint mission between NASA, ESA and the Canadian Space Agency (CSA). NIRSpec was designed and built for the European Space Agency (ESA) by Airbus Defence and Space GmbH in Ottobrunn (Germany), with the two detectors and the Micro Shutter Array provided by NASA Goddard Space Flight Center.


\bibliographystyle{spiebib} 

\end{document}